# Councils in Action: Automating the Curation of Municipal Governance Data for Research

**Maxfield Brown, Eva**

**Weber, Nicholas**

University of Washington | jmxbrown@uw.edu

University of Washington | nmweber@uw.edu

## ABSTRACT

Large scale comparative research into municipal governance is often prohibitively difficult due to a lack of high-quality data. Recent advances in speech-to-text algorithms and natural language processing techniques has made it possible to more easily collect and analyze this type of data. In this paper, we introduce an open-source platform, the Council Data Project (CDP), to curate novel datasets for research into municipal governance. The contribution of this work is two-fold: 1. We demonstrate that CDP, as an infrastructure, can be used to assemble reliable comparative data across municipalities; 2. We provide exploratory analysis to show how CDP data can be used to gain insight into how municipal governments perform over time. We conclude by describing future directions for research on and with CDP such as the development of machine learning models for speaker annotation, outline generation, and named entity recognition to the linking of data for large-scale comparative research.

## KEYWORDS

public interest technology; municipal governance; data archival; computational data access; natural language processing

## INTRODUCTION

Federalism, where power devolves to states and municipalities, is a defining feature of US democracy. But, federalism also poses substantial challenges to the study of state and local governments - each state and each city or county in the USA has their own rules and regulations that structure a mode of governance. For this reason, political science research into local government is often referred to as a "black hole" (Sapotichne et al., 2007). It can be an extended exercise in data collection to determine the identity of elected officials across the 89,000 local governments in the USA, let alone measure and compare the performance of one system of local governance to another (Sumner et al., 2020).

Despite this challenge, there are common legal requirements for transparency in state and local government legislatures that can improve the quality, scale, and usefulness of data used to study local governance. Open meeting laws, also called sunshine laws, require that most meetings of state and local governments be open to the public, along with their decisions and records (Aichinger, 2009). This means, in theory, recordings of a legislative session, bills and materials supporting the proposal of a bill, as well as voting outcomes are required to be publicly accessible. In practice, access to recordings and voting records is often hampered by closed, proprietary information systems that are difficult to navigate, search, and extract meaningful data for both citizens and researchers alike (Ferreira & Gyourko, 2014).

Over the last two years, the Council Data Project (CDP) has developed an open-source software platform that significantly improves access and engagement with local government data (Brown et al., 2021). In the following paper we provide a brief background of CDP, and how this open-source platform enables large-scale, longitudinal, comparative research into municipal governance. In doing so, we position our work amongst other large-scale, public interest technology projects, and describe how CDP uniquely enables the assembly of comparative data for analysis. We then use CDP to assemble a longitudinal dataset from three different municipal councils (Seattle City Council, Portland City Council, and King County Council). We use this dataset to perform exploratory data analysis through the development of an N-gram plotting tool. Finally, we conclude by describing our own future work and the future work made possible by CDP and a growing data collection that we refer to as "Councils in Action."

## BACKGROUND

### Public Interest Technology for Municipal Events

A number of previous civic technology applications have created valuable and accessible local government data from publicly available information, but none have specifically focused on aggregating transcripts of legislative discussion. In the following section we review four of these applications, and note how they relate to but differ from CDP. *Councilmatic* (https://github.com/codeforamerica/councilmatic) was one of the first public interest technology projects to focus on making council information more accessible. Councilmatic is a system for processing and archiving past municipal council meetings and legislative information and tracking upcoming meetings.

Councilmatic is entirely open-source and there are now working examples of this application being used in cities throughout the USA, including for Los Angeles, Chicago, New York City, and Oakland (Poe & Gregg, 2015). However, each instance of Councilmatic, at a city or county level, is entirely separate from each other in their setup and maintenance. This distributed architecture makes it difficult to collaboratively develop new features, and prohibits cross-municipality data aggregation.

*Local Voices Network* (LVN) (https://lvn.org/), a project from Cortico AI, provides a powerful platform for generating, visualizing, and searching through transcripts of civic "conversations." LVN tools are highly targeted at reaching out to communities, facilitating a small-group conversation, and making such conversations easy to digest via machine learning analysis and visualization (Roy, 2020). Such facilitation of community conversations engineers a discussion about specific topics and places them into a standardized format rather than curating existing data into a standard form. LVN produces novel insight into community sentiment but at the high cost of facilitation and engineering.

*Big Local News* (BLN) (https://github.com/biglocalnews), from researchers at Stanford University has created a platform to obtain data, through web-scraping, from municipal councils across the country. Each scraper from BLN collects meeting assets such as documents, presentations, videos, captions, etc.. Further, there is currently no processing of the scraped data (e.g. transcript generation) in order to expand the usage of these documents to include council discussions for analysis.

*Blockparty* (https://blockparty.studio) is an emergent civic technology project which generates and analyzes meeting transcripts from New York City community council meetings. Blockparty creates and processes meeting transcripts to extract keywords and other potential highlights, and then publishes both the transcript and a keyword histogram to the web. Blockparty currently only serves New York City and without open-source code we are limited in understanding their deployment and data access mechanisms. Specifically, we do not know how researchers and civic hackers alike might deploy Blockparty for their own municipality, and more importantly, there is no structured open data produced by this project that can be analyzed for research purposes.

Lastly, while many projects are focused on the collection and republishing of municipal meeting data (from documents, videos, and in some cases transcripts), it is rare for a project to address the data collection, aggregation, transformation, and linkage of data together. For example, Councilmatic is the only project from prior examples which additionally stores legislative outcomes and voting records. The storage of such data is a valuable attachment which allows for investigation not only from discussion but additionally against the legislative end result.

**Council Data Project**

Council Data Project (CDP) attempts to improve upon state-of-the-art public interest technology projects by providing a low-cost, flexible and scalable open-source solution for generating a large standardized corpora of municipal council meetings. CDP can be deployed in any municipality with minimal configuration by a developer or city IT department. Using a simple Python-based 'cookie-cutter template' (Cookiecutter, 2013) a developer can configure a new CDP deployment with just a two-line installation process and fully deploy the instance once provided a function to gather events (Brown et al., 2021). Once fully deployed, an instance of CDP will collect and process municipal meeting minutes, agendas, voting records, and crucially, the recorded video or audio of each meeting that is archived by a municipality. For every legislative meeting that CDP processes, the system generates a transcript from the provided video using a machine learning model that converts recorded audio to text (aka speech-to-text processing). On a continuous schedule CDP infrastructure then uses this corpus of transcripts to generate and update a keyword based index to enable search of meetings by keyword (Brown et al., 2021).

To make CDP data more accessible to the public, and to provide an interface for building a deeper, contextual understanding, each CDP infrastructure also publishes a website (see Figure 1 below). Users of CDP websites can search for municipal meetings using the keyword index and then retrieve the video of a meeting, the sentence time-stamped transcript (for reading and for jumping playback of video to a sentence start point), access the full minutes of the meeting, and view votes that took place during a meeting. Further, because CDP transforms municipal government data into a database specification, CDP also aggregates and publishes aggregate data - such as the entire voting records of city council members, or document timestamps that show when specific actions were taken on a piece of municipal legislation.

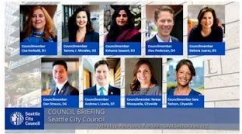 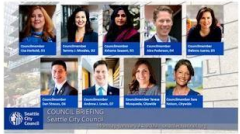 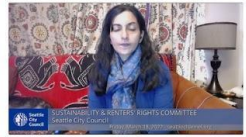
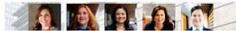 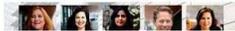 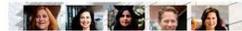

**Figure 1.** Event search results for query: "missing middle housing" using the Seattle City Council CDP instance website. The search page includes event 'cards' which contain a thumbnail from the meeting video, the meeting date, the meeting committee name, a snippet from the meeting transcript that contains and highlights one or more of the keyword search terms which were found in the transcript, and the meeting keywords. Above the event cards, are filters for the returned query for filtering by committee and date. Additionally there are options for sorting the returned results. The Seattle CDP instance website is available at: https://councildataproject.org/seattle/

CDP imposes minimal requirements as to the level of basic information that must be collected for a municipal event to be identified, accessed, and processed. At a bare minimum, for CDP infrastructures to process and store an event, the system must be given a URL to a video of the meeting, the date of the meeting, and the name of the meeting committee (i.e. "Full Council", "Transportation Committee", etc.). This allows CDP instances to be deployed for less resource-available councils (i.e. school boards, neighborhood zoning boards, etc.) while still producing a standardized transcript and access mechanism to both view and download the data for further exploration and analysis. Data ingestion can be customized for each CDP deployment, but the core processing pipelines, infrastructure configuration, and web application are all shared across any city that deploys CDP. This allows for a much easier and larger collaborative effort between developers and open-source software contributors. In the following section, we detail previous text-as-data datasets from government sources that have been constructed and how such CDP deployments can be utilized to compile a large corpus of municipal meeting transcripts for analysis.

## CDP - COUNCILS IN ACTION DATASET

### Previous Government Meeting Datasets

Municipal meeting data is used across a number of domains of research that are interested in the institutional design and functioning of local governments - from political science and sociology to legal scholarship. In the following section we highlight a study from Einstein et al. which investigated who participates in local meetings and a study from Jacobi and Schweers which investigated how gender, ideology, and seniority affect Supreme Court oral argument. Both studies relied upon utilizing meeting records (video or transcript) for analysis of participant behaviors.

Einstein et al. provided a comprehensive look into who participates in local government. Specifically, they "[compiled and coded] new data on all citizen participants in planning and zoning board meetings dealing with the construction of multiple housing units in 97 Massachusetts cities and towns." The researchers then matched, "thousands of individual participants to the Massachusetts voter file to explore who participates in local political meetings" (Einstein et al., 2019). This paper utilized text annotation and topic and sentiment encoding to first identify participants, and then determine what each participant did and did not support in regards to specific planning and zoning discussions. In their study, data collection and coding were done in a combined manual and automated process. Public comment coding and annotation was completed by identifying participants' names and addresses when they spoke. Once the data had been manually collected, Einstein et al. used probabilistic name and address matching with Massachusetts voting records in order to match each participant to their voting record details and then manually verified matches.

In a similar study which used mixed manual and automated methods of constructing a dataset for federal governance research, Jacobi and Schweers attempted to measure the effect of gender, ideology, and seniority at Supreme Court oral arguments. Their work processed hundreds of transcripts to search and record interruptions between the legal advocates and the Supreme Court Justices (and between Justices) (Jacobi & Schweers, 2017). Jacobi and Schweers work was made possible by two separate databases: an existing publicly available database of specifically Roberts Court oral arguments and a second database that was manually assembled to store in-depth analysis of interruption behaviors.

These two examples illustrate how transcript data from governance deliberations can be used to study an enormous range of consequential topics - from gendered speech patterns, to representative democratic outcomes. While these results are individually impactful, the ability to build and expand upon this research is limited because of expensive and time consuming processes required for manually collecting, processing, and structuring data for analysis. In the following sections we describe the content and structure of data made available by CDP instances and how we can make analyses of municipal governance both accessible and reproducible for research. In particular, we detail the construction of a dataset, Councils in Action, and describe how it was prepared as a corpus of machine readable transcripts ready for analysis. We then perform exploratory analysis to demonstrate the value of this corpus for municipal governance research.

### CDP - Councils In Action

Using Council Data Project infrastructures we assemble longitudinal data from across multiple municipal councils to ease manual curation for researchers. The proof-of-concept dataset, Councils in Action, is a corpus of over 350 meetings of the city councils of Seattle Washington and Portland Oregon and the county council of King County Washington. Each meeting in our dataset includes a video file, an audio file, a transcript, and the full meeting minutes (legislative items, votes, and attached documents). Table 1 provides specific details as to the number of meetings from each municipal council and their first and last event dates.

| Instance | Events | First Event | Last Event |
| --- | --- | --- | --- |
| cdp-seattle-21723dcf | 256 | 2021-01-04 | 2022-03-29 |
| cdp-king-county-b656c71b | 72 | 2021-10-05 | 2022-03-30 |
| cdp-portland-d2bbda97 | 32 | 2021-07-07 | 2022-03-30 |

**Table 1. The number of meetings and first and last meeting dates provided by each council in the Councils in Action dataset to date.**

As described in the Council Data Project section, each CDP instance has a website to search, discover, and link data together for a single event. To serve researchers, we further make this dataset available via Python API and ZIP archive download. We provide the *cdp-data* Python library specifically to access, download, cache, and analyze the Councils in Action dataset. For full documentation of all functionality available in the *cdp-data* library please see the provided package documentation: https://councildataproject.org/cdp-data/. For lower-level, direct database access, we provide the Python library *cdp-backend*. More information on lower level access to each instance is made available on each CDP deployment's repository README (i.e. https://github.com/CouncilDataProject/seattle) and extensive documentation as to the CDP database schema is made available via the *cdp-backend* package documentation: https://councildataproject.org/cdp-backend/.

The flexibility in data collection afforded by CDP's distributed instance deployment model allows the dataset to rapidly scale both vertically (in the number of meetings for any single council) and horizontally (as more CDP deployments are created). Therefore, as more CDP instances are created, the generated Councils in Action dataset removes barriers to research that have previously been hindered by time-consuming manual data collection and analysis.

### EXPLORATORY DATA ANALYSIS

In the following section we use the Councils in Action dataset to explore and examine trends in council meetings, including public comments, over time. Our exploratory analysis focuses on keywords or N-grams. N-gram viewers have been commonly created to visualize trends in the usage of specific n-grams in large literature corpora over time (Michel et al., 2011). Such approaches are often considered a way to 'distantly read' a corpus of texts (Organisciak et al., 2022). Distant readings of council meetings can help understand broad trends in the way that a topic increases or decreases in importance during legislative processes. For example, if a topic decreases in frequency then, broadly, we can interpret this topic as being less important in the municipal government's legislative agenda.

For the Councils in Action dataset we apply the use of an n-gram visualization in order to demonstrate how topic trends evolve over time. First we use longitudinal data from the transcripts of Seattle City Council meetings to show the usage of specific n-grams as a percent of total n-grams used for each meeting during this time period. Figure 2 shows the usage of n-grams stemming from "police", "housing", "union", and "homelessness" from January 1, 2021 to April 1, 2022 during meetings of the Seattle City Council.

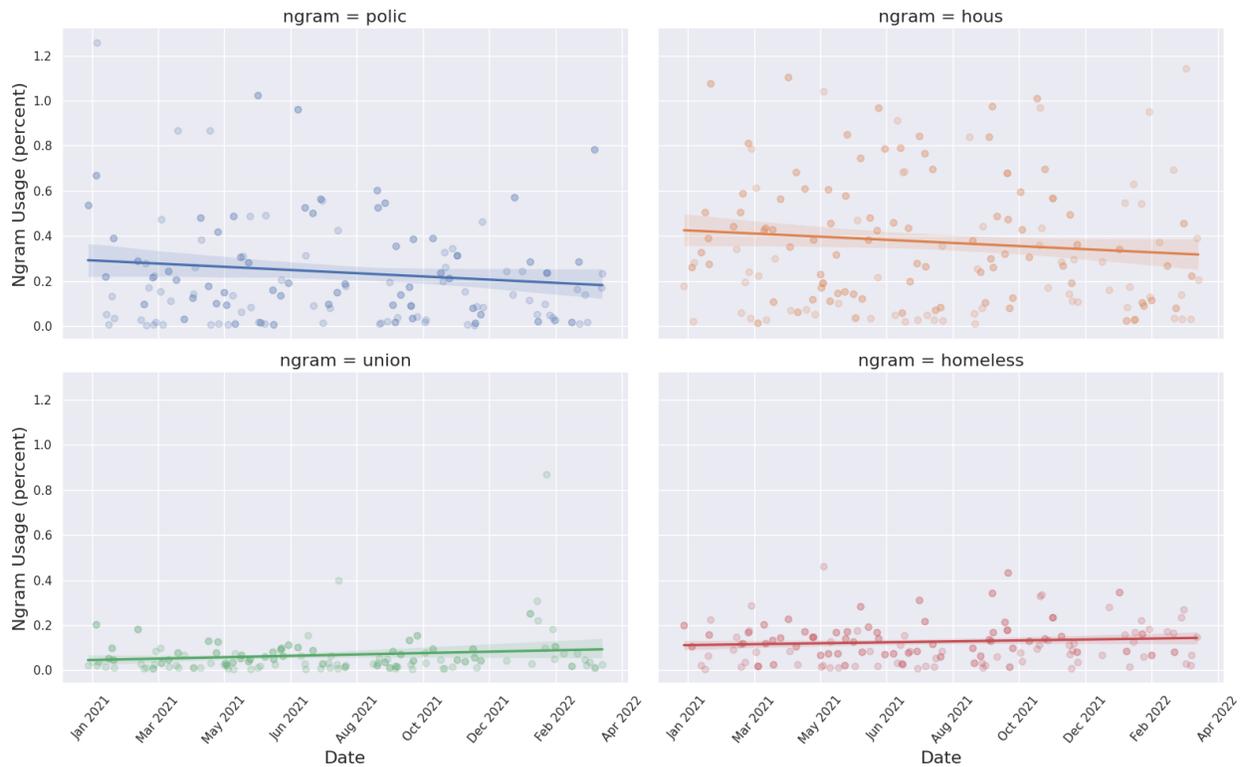

**Figure 2. N-gram usage over time for Seattle City Council meetings from January 1, 2021 to April 1, 2022. The selected n-grams are "polic" (the stem of police, policing, etc.), "hous" (the stem of house, housing, etc.), "union", and "homeless". The y-axis represents the usage of each n-gram – the percent of the number of times the specific n-gram was used for the day over the total number of n-grams used during the day.**

To broaden our n-gram usage counting criteria to more than just our specific query grams, we stem all grams in the dataset using a Snowball stemmer to collect and plot the stemmed n-grams (Bird, 2009). This stemming helps collect and separate words together, for example, "police" and "policing" both stem from "polic" but "policy" stems from "polici" (Porter, 2009). Figure 2 shows that on average, usage of grams stemming from "polic" ("police", "policing", etc.) has declined – from an average usage of 0.39% in January 2021 to an average usage 0.24% in March 2022. Similarly, usage of grams stemming from "hous" ("house", "housing", "houses", etc.) has also declined during the same time period (0.41% in January 2021 to 0.30% in March 2022), but the average usage of such grams remains higher than grams stemming from "polic." There is less overall usage of grams stemming from "union" and "homeless" throughout the entire time period, but these two n-grams do show a slight increase in use.

From a distanced reading approach, we can then infer that over time certain topics became less relevant to the overall debate and legislative agenda of Seattle's City Council. We cannot extend this broad inference to make a causal claim about what topics this particular council cares about or spends time on, but we can show overall trends in what makes up the terms of a legislative session. Such inferences can be valuable in tracking, longitudinally, how a council addresses specific issues, or even how public commenters seek to influence debate and topical relevance. With additional data collected from CDP, we might even be able to show how an event, like the passing of a legislative bill, impacts the frequency of a term over time. For example, after passing a resolution to express support for unionization efforts in Seattle area Starbucks does the city council continue to debate this topic, or does this event cause the topic of unionization to decrease in future meetings? This type of hypothesis can be pursued within a council as the Councils in Action dataset expands over time.

We can further use this same n-gram viewing function to compare across all councils in the Councils in Action dataset to understand if the trends in Seattle hold across the entire dataset. Using the six months of data available for all municipal councils (October 1, 2021 to April 1, 2022), we can investigate if there are any collective trends in n-gram usage. Figure 3 shows the usage of n-grams stemming from "police", "housing", "union", and "homelessness" from October 1, 2021 to April 1, 2022 during meetings of the Seattle City Council, King County Council, and Portland City Council.

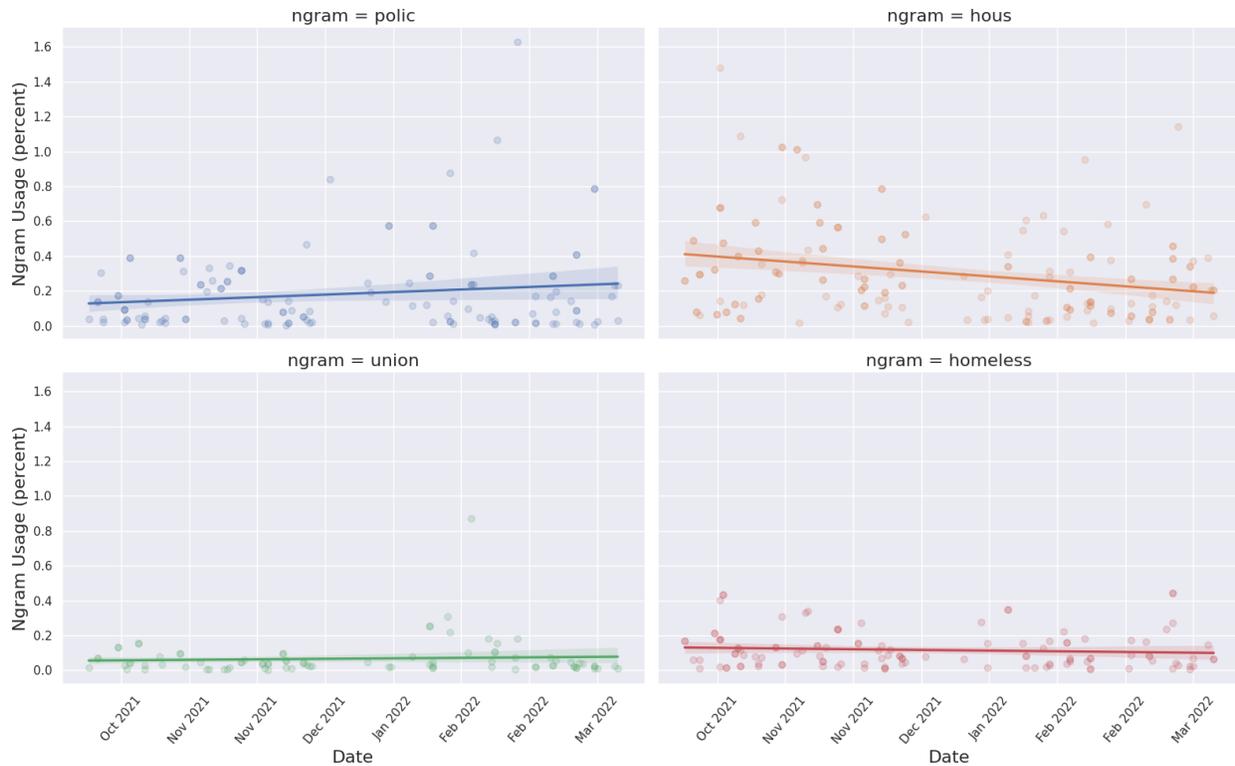

**Figure 3. N-gram usage over time for the entire Councils in Action dataset (Seattle City Council, King County Council, and Portland City Council) from October 1, 2021 to April 1, 2022.**

During this smaller six month timespan we observe some similar trends to the larger time frame Seattle-only data, such as usage of the grams "union" and "homeless" coming up much less frequently than "polic" and "hous", and that the usage of "hous" is declining (from 0.34% in October 2021 to 0.23% in March 2022). However, it seems that, taken as a whole, for this six month period, discussion using grams stemming from "polic" is on the rise, from an average usage of 0.10% in October 2021 to an average usage of 0.19% in March 2022.

For a better understanding of how each individual municipal council is contributing to these overall trends we can plot this same usage data but split out by each municipal council. Figure 4 shows the usage of the same n-grams for the same six month time period during meetings for each of the Seattle City Council, King County Council, and Portland City Council.

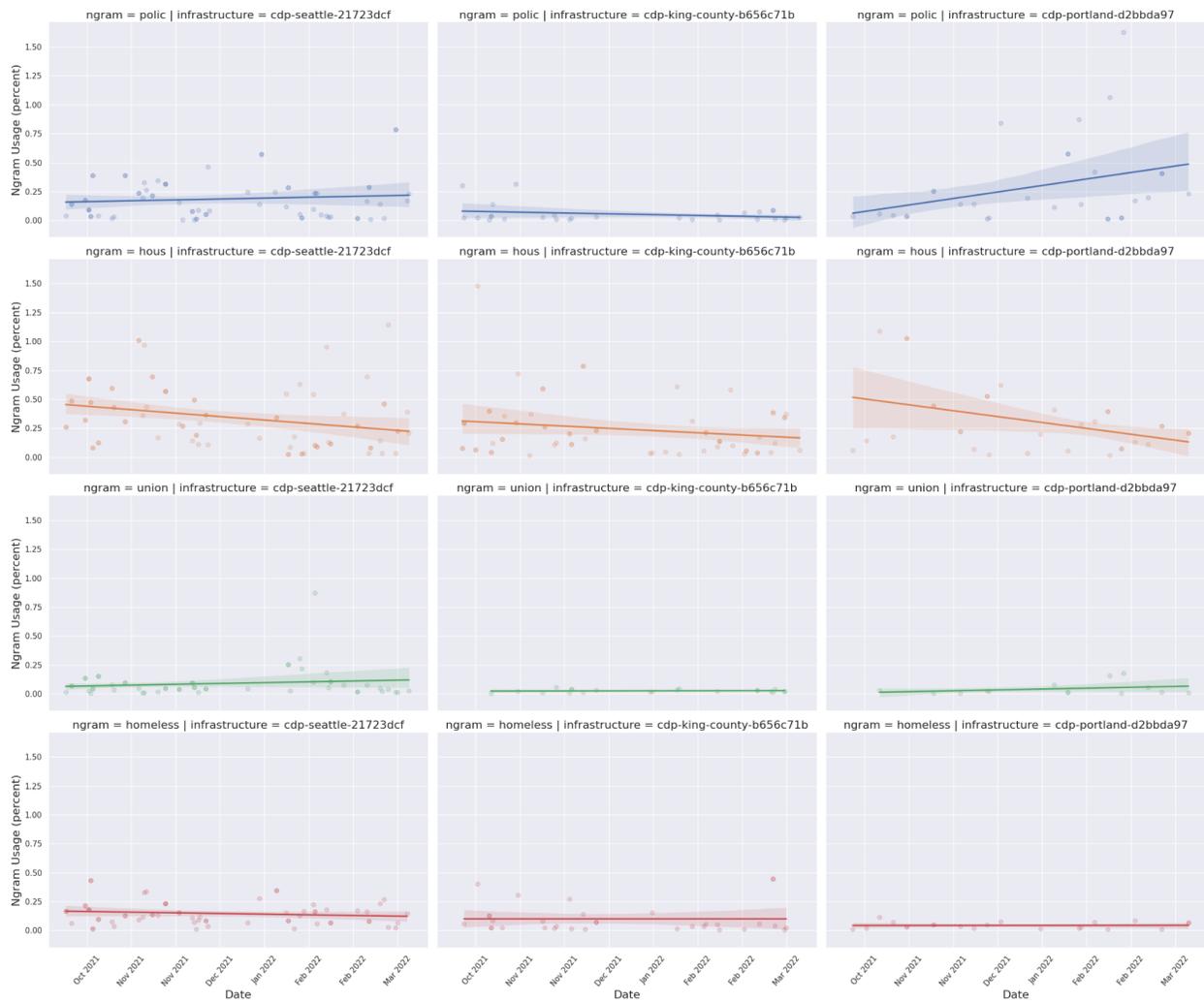

**Figure 4. N-gram usage over time for the entire Councils in Action dataset from October 1, 2021 to April 1, 2022. Each row of plots is a different n-gram and each column of plots is a different CDP instance.**

Much like figure 3, figure 4 shows similar and shared trends for n-grams stemming from "hous", "union", and "homeless", however there are stark differences between the municipal councils in their usage of n-grams stemming from "polic," with Portland City Council's use of n-grams stemming from "polic" rising from an average use of 0.05% in October 2021 to 0.28% in March 2022. We argue that this type of analysis is a powerful tool for understanding local and even regional trends as a whole, but demonstrate that this can also provide insight as to how local events affect council discussion in specific municipalities.

Tracking and following trends of discussion is an important step in understanding what topics public commenters and elected officials are discussing and potentially crafting legislative action for. We want to emphasize that viewing trends over time should not be used as a predictor of future discussion of political action but rather to highlight how CDP and the Councils in Action dataset make it possible to analyze municipal council discussion over time. Further, these observations in discussion trends should not be used as a direct proxy for the amount of legislative action a council is undertaking on a specific issue. Discussion trends are incredibly useful in understanding potentially what public commenters are bringing to council, the seasonality of topical discussion, and topical diffusion and spread across multiple councils but they are just one part of the municipal legislative process. To connect this to previous work, we note that Einstein et al. showed that by measuring and quantitatively investigating municipal board discussions, it is possible to understand who participates in council deliberation and allows researchers to understand trends in sentiment about highly political topics. The CDP Councils in Action dataset enables such work to be done on an even larger scale and potentially affords researchers the ability to answer their questions not only for a single state but for the all municipal councils CDP serves.

## Conclusion

In this paper, we have argued that the deployment of Council Data Project infrastructures to cover municipal councils is a solution to not only increasing access to data, but standardizing this data for eased analysis. We have demonstrated that, with the proof-of-concept Councils in Action dataset, data produced by CDP infrastructures can be easily processed and analyzed to observe shared and unique discussion trends across municipal councils. As the number of CDP instances increases, the Councils in Action dataset can be used for even more rich and varied analyses. For example, in their comprehensive study detailing who participates in local government meetings, Einstein et al., concluded that while there may be suggestive evidence that the trends they found hold for other states, the largest limitation of their work is that the data comes from a single state (Einstein et al., 2019). While the findings of such work cannot be automated, the laborious annotation process required before research can be made easier with models to automatically annotate topical discussion, named entities and the linkage between discussion and legislative action, and the annotation of speaker turns. Because all CDP instances share common processing pipelines, delivering new features to each instance (municipality) that CDP covers is made simple. For example, to replicate a study like Jacobi and Schweers "Justice, Interrupted", but for every municipality covered by CDP, we have already begun to work on an audio-based speaker classification model to label each sentence in a transcript with the known speaker's name and using speaker diarization for labeling each of the unknown speaker's during each meeting (Bredin et al., 2020).

CDP and the Councils in Action dataset can also potentially be used to measure and automatically track the provenance and discussion from legislative action from "model bills" across the country (Alecexposed.org, 2022). A more general form of such work might look to measure the topical and legislative diffusion across the country, for example answering the question: "how long does it take for similar legislative actions regarding a topic to occur in multiple different municipalities?" There are additional computational research questions available for investigation with the Councils in Actions dataset such as the research and development of methods for minutes items and transcript alignment or even more generally, models for "outline generation" to automatically generate the minutes items of a meeting from a transcript (Tardy et al., 2020, Zhang et al., 2019).

Lastly, we emphasize that our proof-of-concept work in this paper demonstrates the possibility of research produced using the Councils in Actions dataset to make its way back into the CDP instances themselves to remove barriers to municipal information for all members of the communities they serve. Council Data Project, as an open infrastructure platform, affords researchers, journalists, activists, and community members the opportunity to directly integrate their work with the data processing pipelines and/or web applications that are connected to CDP deployments. Integration efforts can directly support others working with the Councils in Action dataset, or members of the public hoping to understand the larger context of discussion, track legislative action, and hold elected officials accountable.

## Acknowledgments

We wish to thank the many volunteers that have contributed code, design, conversation, and ideas to the Council Data Project; the University of Washington Information School for support; and, Code for Science and Society and the Digital Infrastructure Incubator for providing guidance on developing a sustainable open-source project.